\documentclass[prl,preprint,showpacs,floatfix,superscriptaddress]{revtex4}
\usepackage{amssymb,graphicx,afterpage,amsmath,color}
\usepackage{verbatim}

\begin{document}
\title{Disorder promotes ferromagnetism: Rounding of the quantum phase transition in Sr$_{1-x}$Ca$_x$RuO$_3$}
\author{L. Demk\'o}
\affiliation{Department of Physics, Budapest University of
Technology and Economics and Condensed Matter Research Group of the Hungarian Academy of Sciences, 1111 Budapest, Hungary}
\affiliation{Multiferroics Project, ERATO, Japan Science and Technology Agency (JST), c/o Department of Applied Physics, University of Tokyo, Tokyo 113-8656, Japan}
\author{S. Bord\'acs}
\affiliation{Department of Physics, Budapest University of
Technology and Economics and Condensed Matter Research Group of the Hungarian Academy of Sciences, 1111 Budapest, Hungary}
\affiliation{Multiferroics Project, ERATO, Japan Science and Technology Agency (JST), c/o Department of Applied Physics, University of Tokyo, Tokyo 113-8656, Japan}
\author{T. Vojta}
\affiliation{Max-Planck-Institut f\"ur Physik Komplexer Systeme, N\"othnitzer Str. 38, 01187 Dresden, Germany}
\affiliation{Department of Physics, Missouri University of Science and Technology, Rolla, MO 65409, USA}
\author{D. Nozadze}
\affiliation{Department of Physics, Missouri University of Science and Technology, Rolla, MO 65409, USA}
\author{F. Hrahsheh}
\affiliation{Department of Physics, Missouri University of Science and Technology, Rolla, MO 65409, USA}
\author{C. Svoboda}
\affiliation{Department of Physics, Missouri University of Science and Technology, Rolla, MO 65409, USA}
\author{B. D\'ora}
\affiliation{Department of Physics, Budapest University of
Technology and Economics and Condensed Matter Research Group of the Hungarian Academy of Sciences, 1111 Budapest, Hungary}
\affiliation{Max-Planck-Institut f\"ur Physik Komplexer Systeme, N\"othnitzer Str. 38, 01187 Dresden, Germany}
\author{H. Yamada}
\affiliation{National Institute of Advanced Industrial Science and Technology (AIST), Tsukuba, Ibaraki 305-8562, Japan}
\author{M. Kawasaki}
\affiliation{Cross-Correlated Materials Research Group (CMRG), RIKEN Advanced Science Institute (ASI), Wako 351-0198, Japan}
\affiliation{WPI-AIMR, Tohoku University, Sendai 980-8577, Japan}
\affiliation{Department of Applied Physics, University of Tokyo, Tokyo, 110-8656, Japan}
\author{Y. Tokura}
\affiliation{Multiferroics Project, ERATO, Japan Science and Technology Agency (JST), c/o Department of Applied Physics, University of Tokyo, Tokyo 113-8656, Japan}
\affiliation{National Institute of Advanced Industrial Science and Technology (AIST), Tsukuba, Ibaraki 305-8562, Japan}
\affiliation{Cross-Correlated Materials Research Group (CMRG), RIKEN Advanced Science Institute (ASI), Wako 351-0198, Japan}
\affiliation{Department of Applied Physics, University of Tokyo, Tokyo, 110-8656, Japan}
\author{I. K\'ezsm\'arki}
\affiliation{Department of Physics, Budapest University of
Technology and Economics and Condensed Matter Research Group of the Hungarian Academy of Sciences, 1111 Budapest, Hungary}

\date{\today}

%%%%%%%%%%%%%%%%%%%%%%%%%%%%%%%%%%%%%%%%%%%%%%%%%%%%%%%%%%%%%%%%%%%%%%%%%%%%%%%%%%%%%%%%%%%%%%%%%%%%%%%%%%%%%%
% Abstract
%%%%%%%%%%%%%%%%%%%%%%%%%%%%%%%%%%%%%%%%%%%%%%%%%%%%%%%%%%%%%%%%%%%%%%%%%%%%%%%%%%%%%%%%%%%%%%%%%%%%%%%%%%%%%%
\begin{abstract}
The subtle interplay of randomness and quantum fluctuations at low temperatures gives rise to a plethora of unconventional phenomena in systems ranging from quantum magnets and correlated electron materials to ultracold atomic gases. Particularly strong disorder effects have been predicted to occur at zero-temperature quantum phase transitions. Here, we demonstrate that the composition-driven ferromagnetic-to-paramagnetic quantum phase transition in Sr$_{1-x}$Ca$_x$RuO$_3$ is completely destroyed by the disorder introduced via the different ionic radii of the randomly distributed Sr and Ca ions. Using a magneto-optical technique, we map the magnetic phase diagram in the composition-temperature space. We find that the ferromagnetic phase is significantly extended by the disorder and develops a pronounced tail over a broad range of the composition $x$. These findings are explained by a microscopic model of smeared quantum phase transitions in itinerant magnets. Moreover, our theoretical study implies that correlated disorder is even more powerful in promoting ferromagnetism than random disorder.
\end{abstract}
\maketitle

%%%%%%%%%%%%%%%%%%%%%%%%%%%%%%%%%%%%%%%%%%%%%%%%%%%%%%%%%%%%%%%%%%%%%%%%%%%%%%%%%%%%%%%%%%%%%%%%%%%%%%%%%%%%%%
% Introduction
%%%%%%%%%%%%%%%%%%%%%%%%%%%%%%%%%%%%%%%%%%%%%%%%%%%%%%%%%%%%%%%%%%%%%%%%%%%%%%%%%%%%%%%%%%%%%%%%%%%%%%%%%%%%%%
Classical or thermal phase transitions generally remain sharp in the presence of disorder, though their critical behavior might be affected by the randomness. On the other hand, zero-temperature quantum phase transitions \cite{SGCS97,Sachdev,LRVW07} -- which are induced by a control parameter such as the pressure, chemical composition or magnetic field -- are  more susceptible to the disorder. Nevertheless, most disordered quantum phase transitions have been found sharp as the correlation length characterizing the spatial fluctuation of the neighboring phases diverges at the transition point.

In recent years, it has become clear that the large spatial regions free of randomness, which are rare in a strongly disordered material and hereafter referred to as \emph{rare regions}, can essentially change the physics of phase transitions \cite{Vojta06}. Close to a magnetic transition, such rare regions can be locally in the magnetically ordered phase
%-- with slow fluctuations leading to the famous Griffiths singularities \cite{Griffiths69} --
even if the bulk system is still nonmagnetic. These rare regions are extremely influential close to quantum phase transitions \cite{Vojta06,Vojta10,GYMBHCHD08,Westerkampetal09,UbaidKassisVojtaSchroeder10}.
% and expected to dominate the thermodynamics. They
%give rise to the the so-called quantum Griffiths phases \cite{Vojta06,Vojta10} as recently observed in magnetic semiconductors \cite{GYMBHCHD08}, heavy-fermion systems \cite{Westerkampetal09}, and transition metal alloys \cite{UbaidKassisVojtaSchroeder10}.

When the rare regions are embedded in a dissipative environment the disorder effects are further enhanced. For example, in metallic magnets, the magnetization fluctuations are coupled to electronic excitations having arbitrarily low energies. This leads to an over-damped fluctuation dynamics. Sufficiently strong damping completely freezes the dynamics of the locally ordered rare regions \cite{MillisMorrSchmalian01},
allowing them to develop a static magnetic order. It has been predicted \cite{Vojta03} that this mechanism destroys the sharp magnetic quantum phase transition in a disordered metal by rounding and a spatially inhomogeneous ferromagnetic phase appears over a broad range of the control parameter.

%%%%%%%%%%%%%%%%%%%%%%%%%%%%%%%%%%%%%%%%%%%%%%%%%%%%%%%%%%%%%%%%%%%%%%%%%%%%%%%%%%%%%%%%%%%%%%%%%%%%%%%%%%%%%%
% Experiments
%%%%%%%%%%%%%%%%%%%%%%%%%%%%%%%%%%%%%%%%%%%%%%%%%%%%%%%%%%%%%%%%%%%%%%%%%%%%%%%%%%%%%%%%%%%%%%%%%%%%%%%%%%%%%% 

The family of perovskite-type ARuO$_3$ ruthanates (with A an alkaline earth ion) offers an ideal setting to test these
predictions. SrRuO$_3$ is a ferromagnetic metal with a Curie temperature of $T_C=165$\,K. On the other hand, no long-range magnetic order develops in CaRuO$_3$ and recent studies indicate paramagnetic behavior or the presence of short-range antiferromagnetic correlations in the ground state \cite{Kolev2002}. It is demonstrated that tiny $Co$ doping can drive the system to a low-temperature spin-glass state \cite{Breard2007}, however, the ground state of CaRuO$_3$ is still under debate. Earlier studies of the transport, thermal and magnetic properties of Sr$_{1-x}$Ca$_x$RuO$_3$ solid solutions revealed that the composition $x$ is an efficient control parameter and the substitution of the Sr ions by the smaller Ca ions gradually suppresses the ferromagnetic character and with it the Curie temperature \cite{scro_suppress_1,scro_suppress_2,scro_suppress_3,scro_suppress_4}. However, estimates of the critical Ca concentration at which $T_C$ vanishes show large variations depending on the way of the assignment, experimental methodology and sample synthesis (e.g. bulk crystals versus thin films with strain due to lattice mismatch with the substrate). In addition, the random distribution of Sr and Ca ions introduces strong disorder in the exchange interactions controlling the magnetic state.

To investigate the magnetic properties of Sr$_{1-x}$Ca$_x$RuO$_3$ with high accuracy, we have grown a
composition-spread epitaxial film of size 10\,mm$\times$4\,mm and thickness 200\,nm ($\sim500$ unit cells) on a SrTiO$_3$ (001) substrate \cite{scro_deposition_1,scro_deposition_2} which sets the easy magnetization direction normal to the film plane \cite{scro_substrate}. The Ca concentration changes linearly from x=0.13 to 0.53 along the long side of the sample, as shown in Fig.~1a. The large atomically-flat area observed in the atomic force microscope image (Fig.~1a) demonstrates the high quality of this film.

The composition and temperature dependence of the magnetic properties of the Sr$_{1-x}$Ca$_x$RuO$_3$ film were
probed by a home-built magneto-optical Kerr microscope equipped with a He-flow optical cryostat. Its
magneto-optical Kerr rotation for visible light is dominated by the charge transfer excitations between the O 2p and Ru 4d t$_{2g}$ states  \cite{scro_moke_mag}. The large magnitude of the magneto-optical Kerr effect, being the consequence of strong spin-orbit coupling in ruthenates \cite{sro_kerr}, was found to be proportional to the magnetization measured by a SQUID magnetometer on uniform thin films. We have performed all these experiments using a red laser diode. The resulting precisions of the magnetization ($M$) and susceptibility ($\chi$) measurements were $6\cdot10^{-3}$\,$\mu_B$ per Ru atom and $8\cdot10^{-3}$\,$\mu_B$T$^{-1}$ per Ru atom, respectively. Since the composition gradient of the sample is about $0.04$\,mm$^{-1}$, the spatial resolution, $\delta\lesssim 20\,\mu$m, of our microscope corresponds to a resolution of $\delta x \approx 0.001$ in the composition,
allowing us to achieve an exceptionally fine mapping of the magnetization versus the control parameter
of the quantum phase transition. See Supplemental Material at [\textit{URL will be inserted by publisher}] for more details on the sample preparation, characterization, and on the experimental methodology.

%%%%%%%%%%%%%%%%%%%%%%%%%%%%%%%%%%%%%%%%%%%%%%%%%%%%%%%%%%%%%%%%%%%%%%%%%%%%%%%%%%%%%%%%%%%%%%%%%%%%%%%%%%%%%%
% Results
%%%%%%%%%%%%%%%%%%%%%%%%%%%%%%%%%%%%%%%%%%%%%%%%%%%%%%%%%%%%%%%%%%%%%%%%%%%%%%%%%%%%%%%%%%%%%%%%%%%%%%%%%%%%%%
An overview of the results is given in Fig.~1b which shows a color contour map of the remanent magnetization $M$ as a function of the temperature $T$ and the composition $x$. It was obtained by interpolating a large collection of $M(x)$ and $M(T)$ curves measured at constant temperatures and concentrations, respectively. The data clearly show that the area of the ferromagnetic phase and the magnitude of the low-temperature magnetization are gradually suppressed with increasing $x$. Figure 2 displays the temperature dependence of the magnetization and susceptibility for selected compositions.
With increasing $x$, the upturn region in the magnetization curves significantly broadens and the width of the ac susceptibility peaks increases. This already hints at an unconventional smearing of the paramagnetic-to-ferromagnetic phase transition at higher values of the composition $x$.
The critical temperature, $T_C(x)$ in Fig.~1b, separating the ferromagnetic and paramagnetic states in the composition-temperature phase diagram was identified with the peak positions in the susceptibility and in the first derivative of magnetization using both the temperature and the concentration sweeps.

The $T_C(x)$ line in Fig.~1b does not show a singular drop at any concentration, instead it grows a tail extending beyond $x=0.52$ where the zero-temperature magnetization is about three orders of magnitude smaller than the saturation value for SrRuO$_3$. Similar behavior is also observed in the low-temperature magnetization $M$ as a function of the composition, $x$, as shown in Fig.~3a. (We found that all $M(x)$ curves measured below T$=$6\,K collapse onto each other without any detectable temperature variation.) $M(x)$ has an inflection point at $x\approx0.44$ followed by a pronounced tail region in which the magnetization decays slowly towards larger $x$. The existence of an ordered ferromagnetic moment is further confirmed by the hysteresis in the $M(B)$ loops
even at $x=0.52$ (see the inset of Fig.~3a). Thus, the evolution of both the magnetization and the critical temperature with $x$ provide strong evidence for the ferromagnetic-to-paramagnetic quantum phase transition being smeared.

%%%%%%%%%%%%%%%%%%%%%%%%%%%%%%%%%%%%%%%%%%%%%%%%%%%%%%%%%%%%%%%%%%%%%%%%%%%%%%%%%%%%%%%%%%%%%%%%%%%%%%%%%%%%%%%%%%%%%%%%%%%%%%%%%%%%%%%%
% Discussion
%%%%%%%%%%%%%%%%%%%%%%%%%%%%%%%%%%%%%%%%%%%%%%%%%%%%%%%%%%%%%%%%%%%%%%%%%%%%%%%%%%%%%%%%%%%%%%%%%%%%%%%%%%%%%%%%%%%%%%%%%%%%%%%%%%%%%%%%
How can the unconventional smearing of the quantum phase transition and the associated tail in the magnetization be understood quantitatively? As the magnetization fluctuations in a metallic ferromagnet are over-damped, sufficiently large Sr-rich rare regions can develop true magnetic order (see Fig.~1c) even if the bulk system is paramagnetic \cite{MillisMorrSchmalian01,Vojta03}. Macroscopic ferromagnetism arises because these rare regions
are weakly coupled by an effective long-range interaction \cite{BK96,BKV05}. To model this situation, we observe that the probability
for finding $N_{\rm Sr}$ strontium and $N_{\rm Ca}$ calcium atoms in a region of $N=N_{\rm Sr}+N_{\rm Ca}$ unit cells (at average composition $x$)
is given by the binomial distribution $P(N_{\rm Sr},N_{\rm Ca}) = \binom{N}{N_{\rm Sr}} (1-x)^{N_{\rm Sr}} x^{N_{\rm Ca}}$.
Such a region orders magnetically if the local calcium concentration $x_{\rm loc}=N_{\rm Ca}/N$ is below some
threshold  $x_c$. Actually, taking finite-size effects into account \cite{Nozadze}, the condition reads $x_{\rm loc} <  x_c - A/L_{RR}^2$
where $L_{RR}$ is the size of the rare region, and A is a non-universal constant. To estimate the total magnetization in the tail
of the transition ($x>x_c$), one can simply integrate the binomial distribution over all rare regions fulfilling this condition.
This yields \cite{Nozadze}, up to
power-law prefactors,
\begin{equation}
M \propto  \exp \left[ -C \frac {(x-x_c)^{2-d/2}}{x(1-x)} \right]
\label{eq:M(x)}
\end{equation}
where $C$ is a non-universal constant. This equation clearly illustrates the notion of ``smeared'' quantum phase transition:
the order parameter vanishes only at $x=1$  and develops a long, exponential tail upon approaching this point.
As $x_c$ represents the composition where the hypothetical homogeneous (clean) system having the average ion size would undergo
the quantum phase transition, the extension of the ferromagnetic phase beyond $x_c$ is an effect of the disorder.
Starting from atomic-scale disorder our theory is applicable as long as a large number of clusters are probed within the experimental resolution, so that the measured quantities represent an average over the random cluster distribution. The smooth dependence of the magnetization on $x$ together with the small spot size of the beam ($<$300\,$\mu$m$^2$) verifies that this is indeed the case. Based on the given spot size the upper bound for the typical cluster size is estimated to be 1-2\,$\mu$m$^2$ (see Supplemental Material).

As a direct test of our theory we fit the lowest-temperature $M(x)$ data with Eq.~(\ref{eq:M(x)}). We take the spatial dimensionality $d=3$ due to the large thickness of the sample far beyond the spin correlation length in the system. As can be discerned in Fig.~3b, the magnetization data in the tail ($x\gtrsim 0.44$) follow the theoretical curve over about 1.5 orders of magnitude down to the resolution limit of the
instrument. For the critical composition of the hypothetical clean system, we obtain $x_c=0.38$, though the quality of the fit is not very sensitive to its precise value because the drop in $M$ occurs over a rather narrow $x$ interval. The composition dependence of the critical temperature $T_C$ can be estimated along the same lines
by comparing the typical interaction energies between the rare regions with the temperature and the same functional dependence on $x$ was found \cite{Nozadze}. The experimental data in the tail region follow this prediction with the same $x_c=0.38$ value, as can be seen from the corresponding fit in Fig.~3b.

Earlier works \cite{scro_suppress_1,scro_suppress_2,scro_suppress_3,scro_suppress_4} on Sr$_{1-x}$Ca$_x$RuO$_3$ resulted in very different estimates (ranging from $x=0.5$ to $1$) for the presumed critical composition of the ferromagnetic quantum phase transition. While this may partially stem from the differences between film
and bulk material, large variations remain present even within one class of samples.
We emphasize that even such large variations can be naturally explained within the smeared
transition scenario as being the result of correlations between the positions
of the Sr and Ca atoms. If the occupation of neighboring unit cells with the same type of
atoms is preferred, the probability for finding large Sr-rich regions will be
greatly enhanced compared to the previous case of independent random occupation of all
cells. Therefore, the tail of the $M(x)$ curve becomes much more pronounced.
To model this situation, we have generalized the mean-field model of Ref.~\onlinecite{Nozadze} to the case of correlated disorder. Figure 4 presents the zero-temperature $M(x)$ curves of several systems with different disorder
correlations ranging from the uncorrelated case to a correlation length
of 2 lattice constants. The figure shows that even short-range correlations
have a dramatic effect; they shift the seeming critical composition
from about $x=0.5$ to $1$. To model and analyze the effects of correlations between the positions of the Sr and Ca atoms, we have considered a (3+1)-dimensional Ising model. The exchange interactions can take two values at random, corresponding to the occupation of the lattice site with a Sr or Ca atom. To generate random variables with the desired spatial correlations, we use the Fourier filtering method \cite{Makse}. The Ising model is then solved within a local mean-field theory which gives the local magnetization on each lattice site.

%%%%%%%%%%%%%%%%%%%%%%%%%%%%%%%%%%%%%%%%%%%%%%%%%%%%%%%%%%%%%%%%%%%%%%%%%%%%%%%%%%%%%%%%%%%%%%%%%%%%%%%%%%%%%%%%%%%%%%%%%%%%%%%%%%%%%%%%
% Summary
%%%%%%%%%%%%%%%%%%%%%%%%%%%%%%%%%%%%%%%%%%%%%%%%%%%%%%%%%%%%%%%%%%%%%%%%%%%%%%%%%%%%%%%%%%%%%%%%%%%%%%%%%%%%%%%%%%%%%%%%%%%%%%%%%%%%%%%%
To summarize, we have studied the paramagnetic-to-ferromagnetic quantum phase transition of Sr$_{1-x}$Ca$_x$RuO$_3$ by means of a composition-spread epitaxial film.
% with $x$ ranging from 0.13 to 0.53.
%Using magneto-optical Kerr microscopy allowed us to achieve an exceptionally fine mapping of the magnetization and susceptibility as functions of  composition $x$ and temperature.
We found that the disorder significantly extends the ferromagnetic phase. Moreover, the phase transition in this itinerant system does not exhibit any of the singularities associated with a quantum critical point. Instead, both the magnetization and critical temperature display pronounced tails towards the paramagnetic phase. The functional forms of these tails agree well with the predictions of our theoretical model.
%which is based on the smeared quantum phase transition scenario \cite{Vojta03}.
Our calculations also show that disorder, if correlated over a few unit cells, is even more powerful in promoting an inhomogeneous ferromagnetic phase. We thus conclude that our results provide, to the best of our knowledge, the first quantitative confirmation of a smeared quantum phase transition in a disordered metal. We expect that this scenario applies to a broad class of itinerant systems with quenched disorder.

%%%%%%%%%%%%%%%%%%%%%%%%%%%%%%%%%%%%%%%%%%%%%%%%%%%%%%%%%%%%%%%%%%%%%%%%%%%%%%%%%%%%%%%%%%%%%%%%%%%%%%%%%%%%%%
% Acknowledgements
%%%%%%%%%%%%%%%%%%%%%%%%%%%%%%%%%%%%%%%%%%%%%%%%%%%%%%%%%%%%%%%%%%%%%%%%%%%%%%%%%%%%%%%%%%%%%%%%%%%%%%%%%%%%%%
We thank A. Halbritter and G. Mih\'aly for fruitful discussions. This work was supported by KAKENHI, MEXT of Japan, by the Japan Society for the Promotion of Science (JSPS) through its ``Funding Program for World-Leading Innovative R\&D on Science and Technology (FIRST Program)'', by Hungarian Research Funds OTKA PD75615, CNK80991, K73361, Bolyai program, T\'AMOP-4.2.1/B-09/1/KMR-2010-0002, as well as by the NSF under grant No.\ DMR-0906566.

\newpage
\begin{figure}[h!]
%\center
%\large{Figure 1}\\
%\vspace{0.2in}
%\includegraphics[width=4in]{Figure1.eps}
\caption{(Color online) Morphology and magnetic characterization of the composition-spread Sr$_{1-x}$Ca$_x$RuO$_3$ epitaxial film. \textbf{(a)} Photographic image of the 10$\times$4\,mm$^2$ film with the local concentration, $x$, indicated along the composition-spread direction. The large terraces of mono-atomic layers in the atomic force microscope image demonstrates the high quality of the film. \textbf{(b)} The contour plot of the remanent magnetization ($M$) over the composition-temperature phase diagram. The dotted mesh is the measured data set used for the interpolation of the surface. The ferromagnetic-paramagnetic phase boundary, $T_C(x)$, derived from the susceptibility and magnetization data (see text for details) is also indicated by the black and grey symbols, respectively. \textbf{(c)} Schematic of the magnetism in the tail of the smeared transition. The spins on Sr-rich rare regions (bright islands) form locally ordered "superspins". Their dynamics freezes due to the coupling to electronic excitations which also tends to align them giving rise to an inhomogeneous long-range ferromagnetic order.}
\end{figure}

\begin{figure}[h!]
%\center
%\large{Figure 2}\\
%\vspace{0.2in}
%\includegraphics[width=2.4in]{Figure2.eps}
\caption{(Color online) Temperature dependence of \textbf{(a)} the remanent magnetization $M$ and \textbf{(b)} ac susceptibility $\chi$ for selected compositions, $x$. The main panel of \textbf{(b)} focuses on the region $x\gtrsim0.4$, and the inset displays representative susceptibility curves over the full range of $x$. Both the magnetization and susceptibility curves show the continuous suppression of the ferromagnetic phase with increasing $x$ and the broadening of the transition.}
\end{figure}

\begin{figure}[h!]
%\center
%\large{Figure 3}\\
%\vspace{0.2in}
%\includegraphics[width=2.7in]{Figure3.eps}
\caption{(Color online) The smearing of the quantum phase transition in Sr$_{1-x}$Ca$_x$RuO$_3$. \textbf{(a)} The composition dependence of the remanent magnetization $M$ at selected temperatures. The inset shows that the hysteresis in the field loops at T=4.2\,K gradually vanishes towards larger $x$ but still present even at $x\approx0.52$. \textbf{(b)} Semilogarithmic plots of the magnetization and the transition temperature $T_C$ as functions of the control parameter in the tail region. The symbols represent the experimental data while solid lines correspond to the theory which predicts $x_c=0.38$ as the location of the quantum phase transition in the (hypothetical) clean system.}
\end{figure}

\begin{figure}[h!]
%\center
%\large{Figure 4}\\
%\vspace{0.2in}
%\includegraphics[width=2.4in]{Figure4.eps}
\caption{(Color online) Model calculation of the magnetization-composition ($M-x$) curves in presence of correlated disorder.
$M_0$ is the saturation magnetization and $\xi$ is the disorder correlation length measured in lattice constants, i.e., it is the length scale over which the occupations of the unit cells with Sr or Ca atoms are correlated.}
\vspace{5in}
\end{figure}

%%%%%%%%%%%%%%%%%%%%%%%%%%%%%%%%%%%%%%%%%%%%%%%%%%%%%%%%%%%%%%%%%%%%%%%%%%%%%%%%%%%%%%%%%%%%%%%%%%%%%%%%%%%%%%
% Figures
%%%%%%%%%%%%%%%%%%%%%%%%%%%%%%%%%%%%%%%%%%%%%%%%%%%%%%%%%%%%%%%%%%%%%%%%%%%%%%%%%%%%%%%%%%%%%%%%%%%%%%%%%%%%%%
\newpage
\begin{figure}[t!]
\center
\large{Figure 1}\\
\vspace{0.4in}
\includegraphics[width=4.4in]{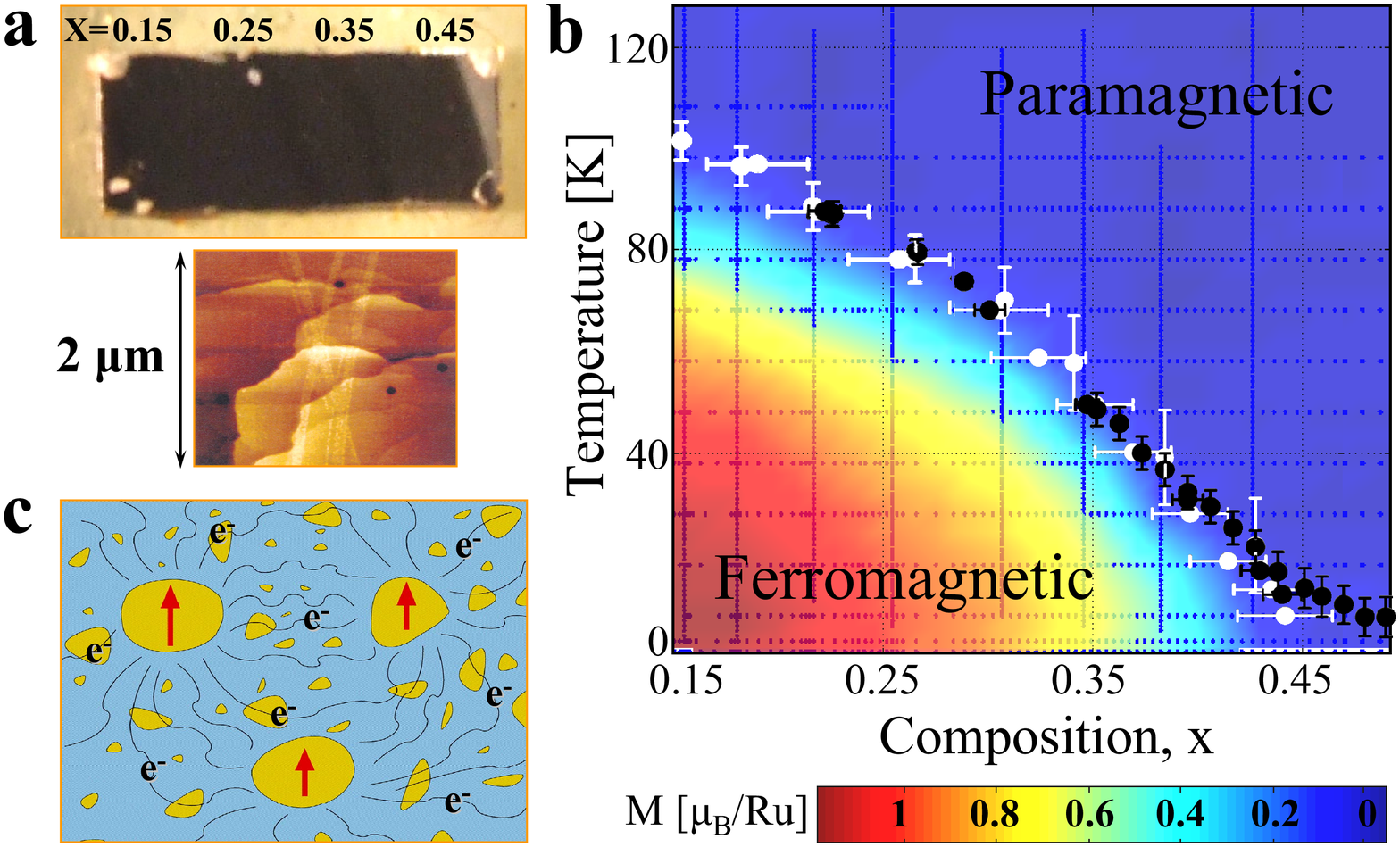}
\end{figure}

\newpage
\begin{figure}[h!]
\center
\large{Figure 2}\\
\vspace{0.4in}
\includegraphics[width=3in]{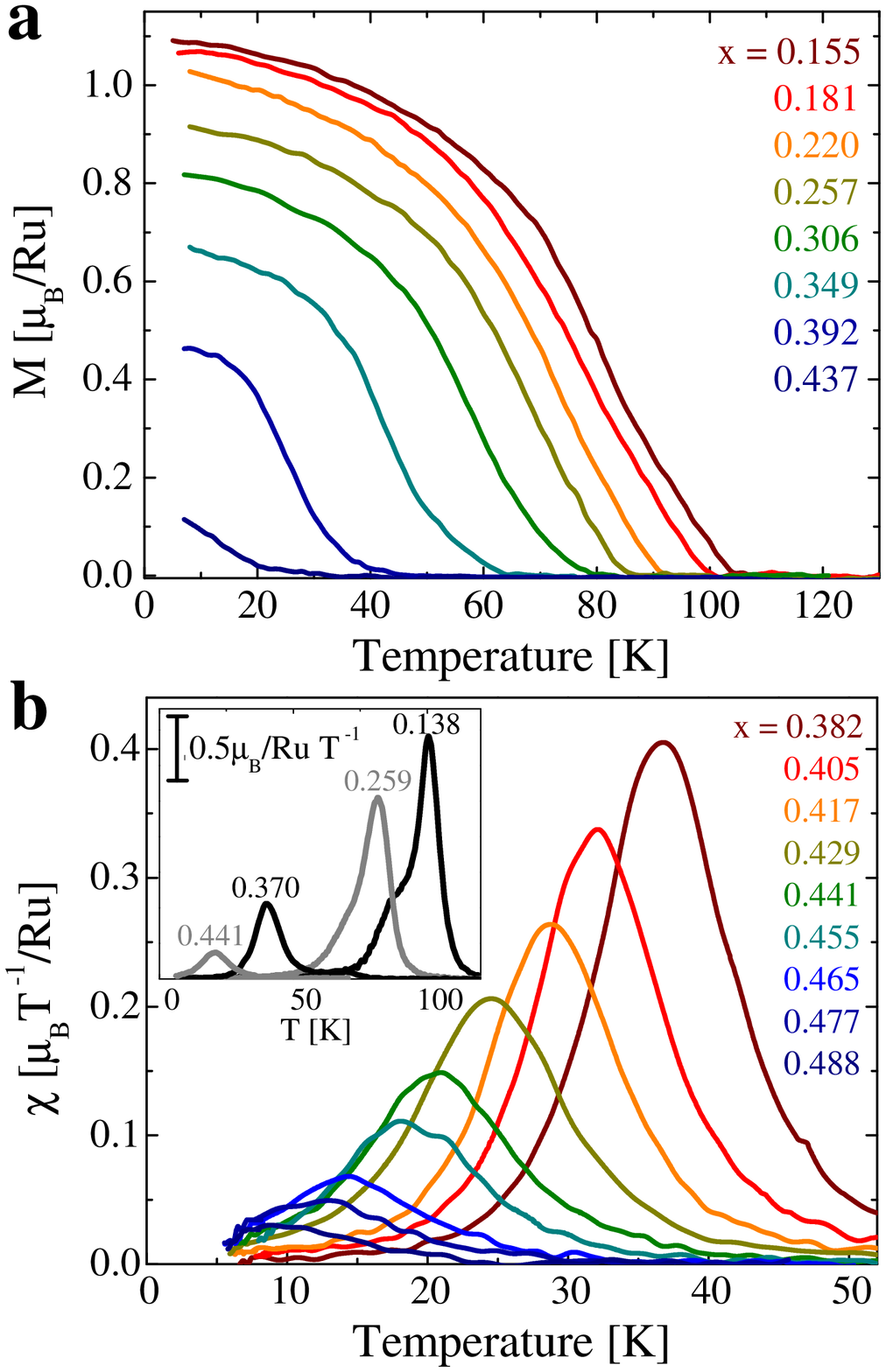}
\vspace{3in}
\end{figure}

\newpage
\begin{figure}[h!]
\center
\large{Figure 3}\\
\vspace{0.4in}
\includegraphics[width=3.2in]{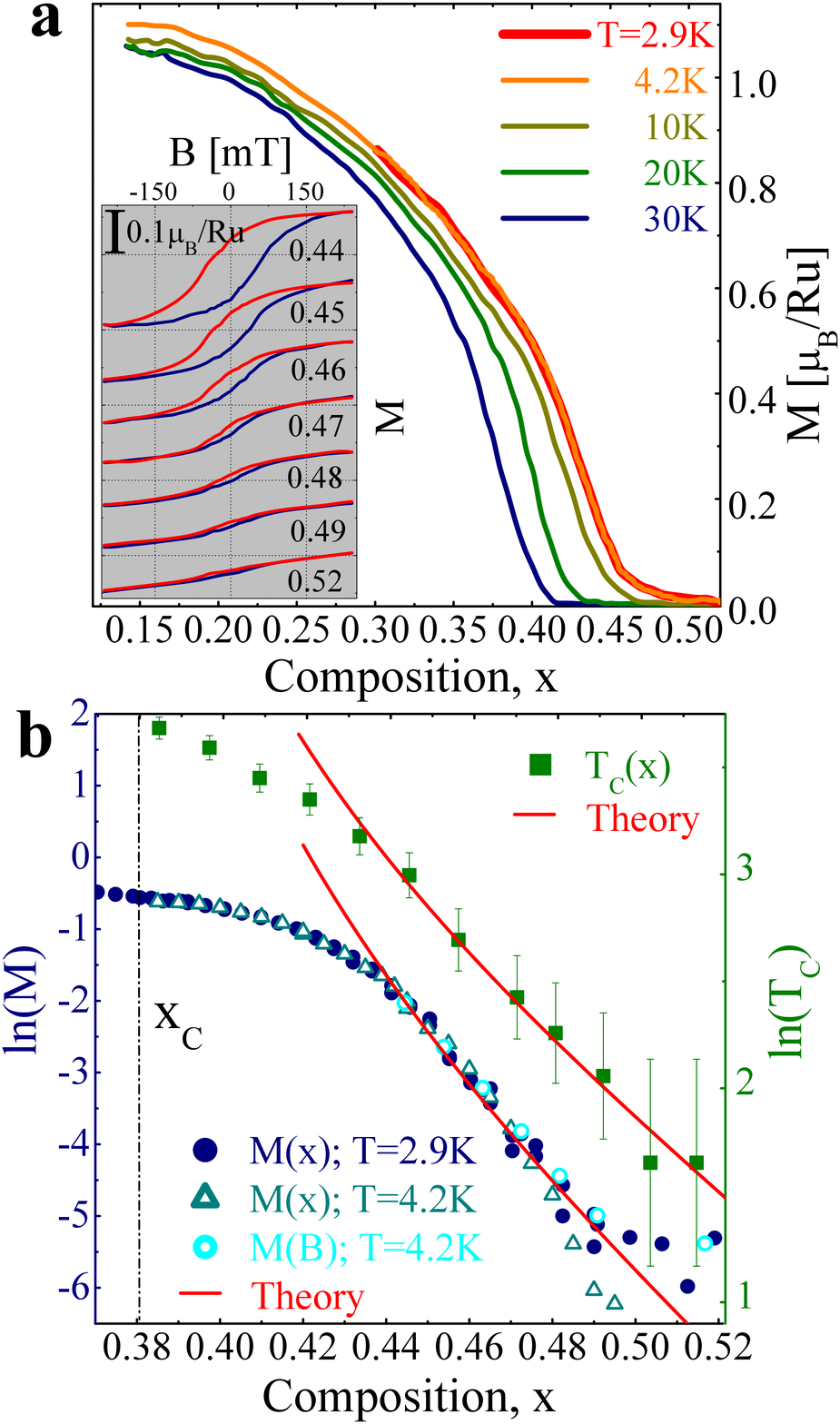}
\vspace{3in}
\end{figure}

\newpage
\begin{figure}[h!]
\center
\large{Figure 4}\\
\vspace{0.4in}
\includegraphics[width=3in]{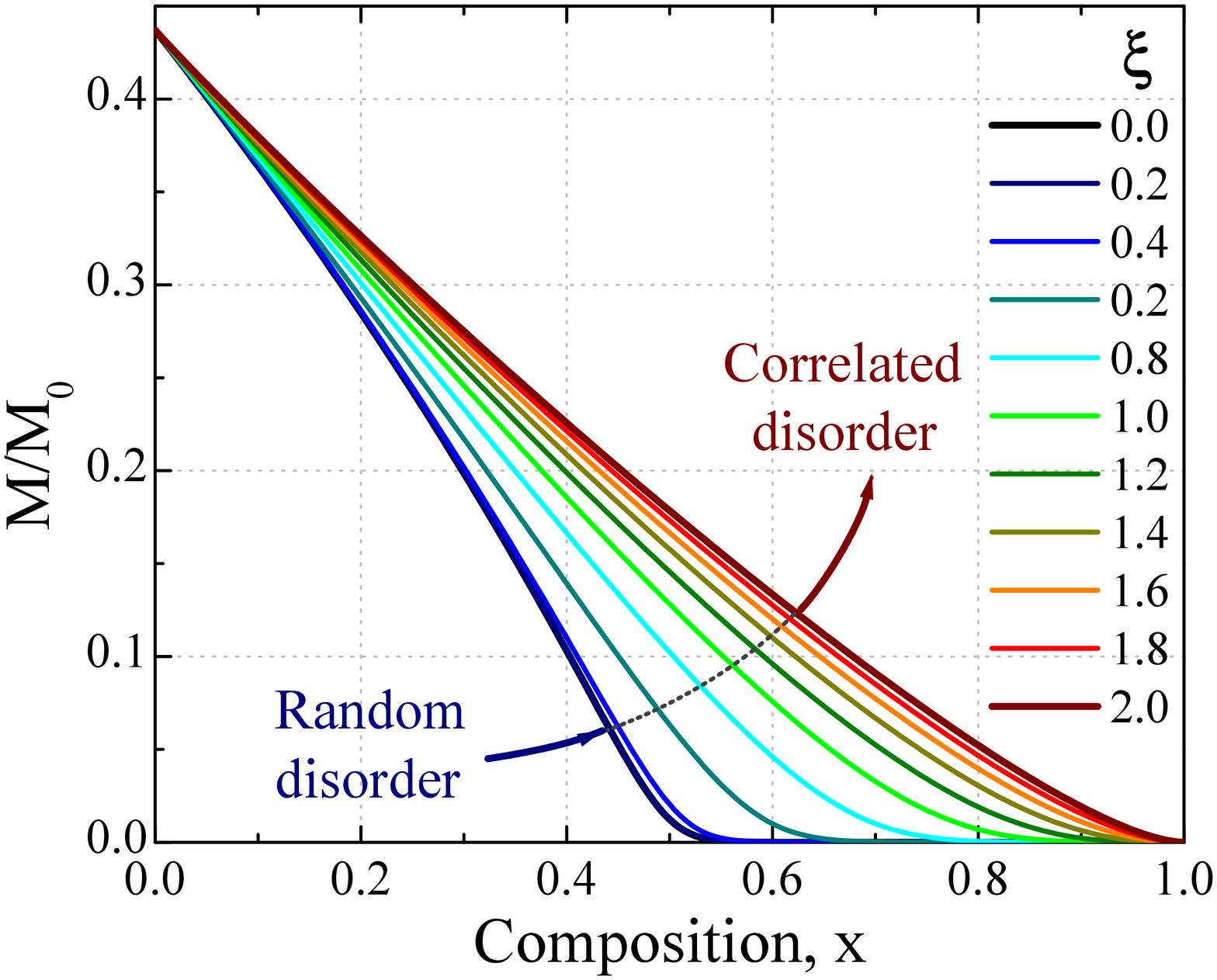}
\end{figure}

\end{document}